\documentclass[prd,twocolumn,preprintnumbers,secnumarabic,amsmath,amssymb,superscriptaddress,nofootinbib,floatfix]{revtex4}
\usepackage{graphicx}% Include figure files
\usepackage{dcolumn}% Align table columns on decimal point
\usepackage{bm}% bold math

\def\beq{\begin{equation}}
\def\eeq{\end{equation}}
\def\be{\begin{equation}}
\def\ee{\end{equation}}
\def\bea{\begin{eqnarray}}
\def\eea{\end{eqnarray}}
\def\d{{\rm d}}

\def\mpl{M_{pl}}
\def\half{\frac{1}{2}}

\newcommand{\GeV}{\mbox{GeV}}

\def\vp{\varphi}
\def\r{\rho}

\def\m{\mu}
\def\n{\nu}

\def\vp{\varphi}

\def\r{\rho}
\def\s{\sigma}

\def\z{\zeta}
\def\a{\alpha}

\def\d{\partial}
\def\gs{g_{\rm s}}

\begin{document}

\preprint{IPMU-09-0009}
\preprint{DCPT-09/09}

\title{Circumventing the eta problem in building an inflationary model in string theory}

\medskip\
\author{Damien A. Easson}%
\email[Email:]{easson@asu.edu}
\affiliation{Department of Physics, and School of Earth and Space Exploration, and Beyond Center, Arizona State University,
Tempe, AZ, 85287-1504, USA}
\affiliation{Centre for Particle Theory, Department of 
Mathematical Sciences, Durham University,
Science Laboratories, South Road, Durham, DH1 3LE, United Kingdom}
\affiliation{Institute for the Physics and Mathematics 
of the Universe, University of Tokyo,
5-1-5 Kashiwanoha, Kashiwa, Chiba 277-8568, Japan}
%%%
\author{Ruth Gregory}
\email[Email:]{r.a.w.gregory@durham.ac.uk}
\affiliation{Centre for Particle Theory, Department of 
Mathematical Sciences, Durham University,
Science Laboratories, South Road, Durham, DH1 3LE, United Kingdom}

\date{\today}
\begin{abstract}
The eta problem is one of the most significant obstacles to 
building a successful inflationary model in string theory. Planck 
mass suppressed 
corrections to the inflaton potential generally lead to 
inflaton masses of order the Hubble scale and generate 
contributions of order unity to the $\eta$ slow roll parameter
rendering prolonged slow roll inflation impossible. We demonstrate 
the severity of this problem in the context of brane anti-brane 
inflation in a warped throat of a Calabi-Yau flux compactification with 
all phenomenologically dangerous moduli stabilized. Using 
numerical solutions we show that the eta problem can be avoided 
in scenarios where the inflaton is non-minimally coupled to gravity and has 
Dirac-Born-Infeld (DBI) kinetic term. We show that
the resulting cosmic microwave background (CMB) observables such as measures of
non-gaussianites can, in principle, serve as a probe of scalar-gravity couplings.

\end{abstract}

\maketitle

\section{Introduction}
It has proven notoriously difficult to connect the inflationary 
universe paradigm \cite{Inflation} with fundamental particle physics. 
However, due to
the recent advances in our understandings of flux compactifications, 
moduli stabilization and various compactification effects 
\cite{Compact,Baumann:2008kq,holosys},  string theory is 
beginning to provide relatively concrete settings for inflationary 
model building~\cite{McASilv}. One of the most severe 
difficulties, preventing
the construction of successful models, is the eta problem. 
For a canonically normalized inflaton $\vp$ of mass $m$, the 
slow roll parameter $\eta$ as a function
of the Hubble paramter $H$ is:
$
\eta =  \mpl^2 \frac{V''}{V} = \frac{m^2}{3 H^2}
\,,
$
where prime denotes differentiation with respect to $\vp$.
Therefore, an inflaton mass of order the Hubble scale is equivalent to 
$\eta \sim \mathcal{O}(1)$. Successful slow roll inflation 
yielding approximately $60$ e-foldings
of inflation requires $m^2 <\!\!\!< H^2$.  In string theory, 
Planck suppressed corrections to the inflaton potential of the form
\be\label{slowk}
\mathcal{\hat O}_4 \frac{\vp^2}{\mpl^2}
\,,
\ee
for some operator of dimension four $\mathcal{\hat O}_4$, generically lead 
to $\eta \sim 1$, when $\left< \mathcal{\hat O}_4 \right> 
\sim V$~\cite{Copeland:1994vg}.  

In this paper, we explicitly show that the eta problem can be 
avoided in models of relativistic brane inflation where the noncanonical 
nature of the inflaton kinetic term
is significant. We present numerical solutions that demonstrate 
this effect, providing concrete realizations of the mechanism envisioned 
by Silverstein and Tong in \cite{AST}. We are primarily concerned with a
contribution to the mass term coming from a non-minimally coupled inflaton
field $\xi \vp^2 R$, although we present a general discussion of problematic mass terms
in Section~\ref{sec:pmt}.
We point out the interesting fact that gravitational couplings in DBI inflation
models alter observable quantities and the sound speed. This can lead to new observational
signatures in the cosmic microwave background (CMB) radiation.

A particularly well understood example of the eta problem 
occurs in brane inflation, \cite{braneinf, KKLMMT}, where the inflaton is 
a (slow-roll canonical) scalar field  $\vp = \sqrt{T_3} r$ 
parameterizing the position $r$ of a D3-brane
moving towards an anti-D3-brane in a warped throat region of a 
Calabi-Yau flux compactification. 
Here, $T_3 = \left((2\pi)^3 g_s \a'^2 \right)^{-1}$ is the 
brane tension and $g_s$ is the string coupling.
Inflation ends when the proper distance between the brane and the anti-brane 
reaches the string length $\ell_s \sim \sqrt{\alpha'}$ and 
tachyon condensation occurs. 
The effective inflaton potential is usually written in the form
\be\label{potef}
V(\vp) = V_{D3/\overline{D3}} + H^2 \vp^2 + \Delta V(\vp)
\,.
\ee
The $V_{D3/\overline{D3}} $ is the attractive Coulomb potential 
between the brane and the anti-brane. This is extremely flat for large $\vp$. 
The second term is the mass term $H^2 \vp^2$ associated with (\ref{slowk})
and generically ruins slow-roll inflation. The mass term is related 
to moduli stabilization effects~\cite{holosys} and was calculated
in \cite{KKLMMT}. The final term $\Delta V$ includes all possible additional  
perturbative and non-perturbative corrections. 

A recent important result~\cite{holosys}
demonstrates that under specific conditions a non-chiral 
dimension two CFT operator  $\mathcal{\hat O}_2$ interacting with
a bulk moduli field $X$ induces 
a negative contribution to the radial potential:
\be
\Delta V(\vp) = - c \,a_0^4 \, T_3  \left( \frac{\vp}{ \vp_{UV}} \right)^2
\,,
\ee
which may be tuned to cancel the problematic second term 
in (\ref{potef}). In the above, $c>0$,  $a_0$ is related to the minimal warp factor in the throat
and $\vp_{UV}$ relates to the ultra-violet (UV) cutoff of the throat geometry~\cite{holosys}. 

While this result is a significant step towards addressing the eta 
problem in the context of the specific model outlined above, we 
will focus on another
possibility. In D-brane inflation models, the inflaton kinetic term is 
of the Dirac-Born-Infeld (DBI) form~\cite{Aharony:1999ti}, 
containing an infinite sum of 
higher derivative kinetic operators. The nonstandard form of the kinetic 
Lagrangian imposes an effective speed limit on the inflaton field allowing 
for a new form of slow-roll inflation even in very 
steep potentials~\cite{AST}. 
Here, we explicitly demonstrate the ability of the DBI 
mechanism to circumvent the eta problem by providing numerical 
solutions giving rise to at least 60 e--foldings of inflation while in the
presence of a conformally coupled mass term~\footnote{An interesting inflationary model 
driven by a non-minimally coupled Standard Model Higgs field appeared in~\cite{Bezrukov:2007ep}. }. 

\section{Compactification Data}
The setting for our scenario is a flux compactification
of Type IIB string theory on an orientifold of a 
Calabi-Yau threefold~\cite{Douglas:2006es}. The line element is
\be
ds^2 = h^{-1/2}(y) g_{\m\n} dx^\m dx^\n + h^{1/2}(y) g_{mn} dy^m dy^n
\,,
\ee
where $h(y)$ is the harmonic function of the manifold, $g_{\m\n}$ is 
the four-dimensional metric and $g_{mn}$ is the metric on the internal space.
We assume the internal space
has a conical throat where the metric may be written locally as
$
g_{mn} dy^m dy^n = dr^2 + r^2 {ds^2}\!\!\!_{X_5}
$,
where $X_5$ is some Sasaki-Einstein space (for example $S^5$ 
or $T^{(1,1)}$ which is topologically $S^2 \times S^3$). 
For sufficient flux, the throat may be strongly warped and this
warping is captured in the form of the harmonic function 
(or ``warp-factor'') $h(y)$. The throat is 
smoothly glued to the compact space and the complex structure 
moduli are fixed as in~\cite{GKP}. The setup is assumed to have 
a single K\"ahler  modulus $\rho$.  The modulus is stabilized via 
strong gauge dynamics on the worldvolume of $n$ supersymmetrically 
wrapped D7-branes on a four-cycle, which generate
a nonperturbative contribution to the superpotential, $W_{\rm np} 
= A(z_\alpha) \exp[{-2\pi \rho/n}]$ \cite{KKLT}.  The prefactor $A$
depends on the three complex D3-brane coordinates 
$z_\a = \{z_1, \, z_2, \, z_3 \}$.
The warped throat is supported by a background D3-brane 
charge $N\gg1$ and is locally well approximated by the 
near horizon D3 brane geometry\footnote{Here
we integrate out the angular degrees of freedom focusing on the radial
motion of the brane as in~\cite{Baumann:2008kq};  however, in general, 
angular structure may play an interesting role \cite{spin}.} 
$AdS_5 \times S^5$, with
warp factor $h$, and radius of curvature, $\Re$, given by:
\be\label{warh}
h(y) = \left(\frac{\Re}{r} \right)^4 \,, 
\qquad \Re^4 = 4\pi  \gs N \alpha'^2
\,.
\ee

In the following discussion it will be useful to define the function 
$f$, related to the harmonic function of the warped compactification 
(\ref{warh}), by, $f(\vp)^{-1} = T_3 h(\vp)^{-1}$. We
also introduce the parameter 
\be\label{eqlam}
\lambda \equiv T_3 \, \Re^4 = \frac{N}{2\pi^2}
\,,
\ee
so that $f(\vp) = \lambda/\vp^4$.

A major difficulty in inflationary model building in the above 
setting is the upper bound on the inflaton field range \cite{Chen:2006hs, BM}:
\be\label{bmbound}
\frac{\Delta \vp}{\mpl} \leq \frac{2}{\sqrt{N}}
\,,
\ee
where this particular expression for the bound was 
derived by  Baumann and McAllister (BM bound) in \cite{BM}.

For the purpose of this demonstration, we consider 
the so-called {\it delicate model} of slow-roll brane 
inflation\footnote{Our results are applicable to more generic 
situations for example the recently constructed models 
of \cite{holosys}.} \cite{Baumann:2008kq}.
This model  hinges on the calculation of 
a one-loop correction to the volume-stabilizing nonperturbative superpotential  
\cite{Compact}, which (for fine-tuned values of the microphysical 
parameters) leads to an effective approximate inflection point 
potential for the inflaton. The potential  is well approximated by
\be\label{inpot}
V(\vp) = V_0 \left( 1 + \lambda_1 (\vp - \vp_\star) 
+\frac{1}{3!} \lambda_2 (\vp - \vp_\star)^3 \right)
\,,
\ee 
where $\vp_\star$ is the location of the inflection point. After sufficient tuning of parameters,
it is possible to construct a viable slow-roll inflationary model\footnote{Recent work suggests 
the overall tunings may be less severe than initially anticipated \cite{Hoi:2008gc, Cline:2009pu}.}.  
We consider this particular form for the potential because inflection
point potentials are a common feature of many string inflation models \cite{Linde:2007jn}, although
our results are applicable to more general potentials.

Our strategy will be to numerically construct a successful model 
of slow-roll brane inflation using the above set-up. We 
then demonstrate that this inflationary solution is lost if we include
the effects of a particular dimension-six Planck suppressed operator 
correction term $\mathcal{\hat O}_6$  of the form (\ref{slowk}). 
Finally, we demonstrate that inflation can be salvaged if one
appropriately takes into account the DBI nature of the inflaton 
kinetic term. Note that we are not concerned with building 
inflationary solutions that conform to the most recent
observations, only in building solutions that give rise to 
more than $\mathcal{N}=60$ e--foldings of inflation and are 
consistent with all known compactification constraints. 
\section{Phenomenology of problematic mass terms}
\label{sec:pmt}
Multiple sources can ultimately contribute to Hubble mass correction terms of
the form $H^2 \vp^2$. We devote this Section to a rudimentary discussion 
of this potentially confusing topic. 
In effective field theory (EFT), one expects a gravitational coupling term, $\xi R \vp^2$,
to appear in the scalar action due to the nature of quantum fields in a curved spacetime.
The term is renormalizable by power counting arguments and therefore
must be included in the curved spacetime scalar field Lagrangian.
Even if it is not present in the bare Lagrangian it will be 
generated by quantum effects: minimal coupling ($\xi =0$) is non-generic
and unstable to quantum corrections. Moreover, the term has been explicitly
calculated in the case of a Dp-brane probe in a negatively curved Einstein
space with conformal boundary with a curved boundary metric (such
as asymptotically $AdS_{p+2}$)~\cite{Seiberg:1999xz,Fotopoulos:2002wy}.
In this case, the case relevant for brane anti-brane inflation, $\xi$ has the 
conformal value\footnote{Note, 
the conformal symmetry is broken by the presence of the
potential and by the DBI nature of the kinetic term.}
$\xi =1/6 $.

During inflation, the Hubble parameter is essentially constant ($\dot H \simeq 0$).
The Ricci scalar in a cosmological background is $R = 6(2H^2 + \dot H)$
so that the conformal coupling term
\be\label{conft}
\frac{1}{12} R \vp^2 \simeq H^2 \vp^2
\,,
\ee
strongly resembles the Hubble scale mass term in (\ref{potef}). Due to this similarity
the source of the mass term was conjectured to be the conformally coupled scalar
in the brane anti-brane inflationary model of~\cite{KKLMMT}. However, this was not
explicitly derived. All that was derived was the pure mass term $H^2 \vp^2$, analogous to the
famed supergravity (SUGRA), dimension-six operator mass term, $\hat{\mathcal{O}}_6 = V_0 \bar \vp \vp$, that results from expanding an F-term potential
and using the Einstein equations in an Friedmann-Robertson-Walker (FRW) background to identify $V_0$ with the
Hubble parameter during inflation $H_0$. An important point, is that the SUGRA mass term is present \it independently \rm from
the conformal term (\ref{conft})\footnote{In addition, specific symmetries may
force $\xi =0$. We thank Andrei Linde for discussions of these points.}.

Besides the conformal term (\ref{conft}), the mass term from the SUGRA analysis \cite{Copeland:1994vg,KKLMMT} and
the terms arising due to moduli-stabilization effects \cite{Compact, Baumann:2008kq, holosys} there are 
other possible EFT sources of problematic mass terms, for example, finite temperature effects. Typically, finite
temperature corrections to the one-loop effective inflaton potential in the high temperature limit
give rise to a temperature dependent mass term of the form  \cite{rhb}:
\be\label{ttime}
\Delta V^T = C T^2 \vp^2
\,.
\ee
Gibbons and Hawking used the path-integral approach to quantum field theory to prove
that Green's functions in de Sitter spacetime are periodic in imaginary time and
yield the Hawking temperature \cite{Gibbons:1977mu}:
\be\label{dstemp}
T_{dS} = \frac{H}{2 \pi}
\,.
\ee
Each observer in de Sitter space has an event horizon, which radiates a thermal spectrum of
particles at the temperature (\ref{dstemp}). Combining (\ref{ttime}) with (\ref{dstemp}) yields a 
finite temperature effective mass term correction\footnote{We thank 
Henry Tye for discussions of this point.} $\sim~H^2 \vp^2$.

Finally, we mention that it is perfectly consistent to include mass terms from both the non-minimally coupled
scalar and various other contributions mentioned above. An interesting example, where the non-minimal coupling term
appears in addition to an explicit  $H^2 \vp^2$ mass term is in the calculation of the Hubble 
effective potential at one loop in~\cite{Janssen:2009pb}.

Ultimately, the origin of problematic mass terms in (\ref{potef}) is 
not important for our results pertaining to a successful inflationary trajectory. Although in general
there are important distinctions between pure mass terms and the conformal coupling term. 
For specificity, we consider a
contribution from the non-minimally coupled
inflaton field parameterizing the D3 brane position as in
\cite{Seiberg:1999xz,KKLMMT}\footnote{We thank Juan Maldacena 
and Shinji Mukohyama for clarifications of this point.}.
The 4D effective action is $S = S_{ST} + S_\vp$, 
where $S_{ST}$ is the scalar-tensor action composed 
of the Einstein-Hilbert action plus non-minimally coupled
scalar mass term
\be\label{actst}
S_{ST} = \int d^{4} \! x \,  \sqrt{-g} \, \left\{ \frac{M_{pl}^2}{2} 
R - \frac{\xi}{2} R \vp^2 \right\}
\,,
\ee
and $S_\vp$ is the (slow-roll limit) canonically normalized 
inflaton kinetic term plus potential $V(\vp)$. In this paper,
we will consider the conformal value $\xi=1/6$ and the $\xi=0$ case
corresponding to minimal coupling. The specific form for
the potential will not play a significant role in our results.

Varying the action with respect to the metric gives:
\bea
(M_{pl}^2 - \xi \vp^2) G_{\m\n}  = \d_\m \vp \, \d_\n \vp -  \half  \d^\r \vp \d_\r \vp \, g_{\m\n}&& \\ \nonumber
- \xi \Big(\nabla_\m \nabla_\n (\vp^2) - \nabla_\s  \nabla^\s (\vp^2) g_{\m\n}\Big)  - V(\vp) g_{\m\n} && .
\eea
The equation of motion for $\vp$ is: 
\be
\nabla^\m  \nabla_\m \vp 
 -  V' -  \xi R \vp =0
\,.
\ee

\section{Numerical Solutions}
\begin{figure}[htbp!]
\centering
\includegraphics[width=0.45\textwidth]{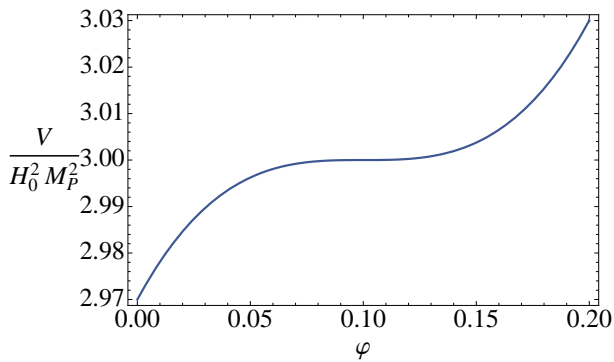}
\caption{Inflection point potential (\ref{inpot}).}
\label{potin.eps}
\end{figure}

For an observationally favored, flat ($k=0$), cosmological (FRW) metric 
ansatz
\be
ds^2 = -dt^2 + a^2(t) d{\bf x}^2
\,,
\ee
the Friedmann equation and equation of motion for $\vp$ are:
\bea\label{nreomn}
3 (M_{pl}^2 - \xi \vp^2) H^2 = \half \dot \vp^2 
+ 6 H \xi \vp \dot \vp  + V(\vp) \,, \\
\ddot \vp + 3 H \dot \vp +  6 \xi \vp (2 H^2 
+ \dot H) + V' =0 \label{nreomp}
\,,
\eea
where $a(t)$ is the scale factor as a function of cosmic time $t$ and $H =\dot a/a$ is the Hubble parameter.
We consider the inflection point potential (\ref{inpot}).
In the following solutions we adopt the parameter values $\lambda_1 
= 7 \times 10^{-5}$, $\lambda_2 = 60$ of \cite{Underwood:2008dh}, chosen to 
produce the correct normalization of density perturbations, 
although in this example we take
$V_0 = 10^{-8}$ corresponding to a GUT inflationary 
scale $M \simeq V_0^{1/4} \simeq 10^{16} \, \GeV$ 
and $\vp_\star = .1$. All quantities are given in Planck units. 
We emphasize that these particular numbers
are not essential to our general findings. The potential 
is plotted in Fig.\ref{potin.eps}.
The throat is supported by $N$ dissolved D3-branes;
in terms of a re-scaled $\lambda$ parameter of (\ref{eqlam}), 
$\bar \lambda \equiv \lambda (H_0/\mpl)^2$ and the inflationary scale $M$:
\be
N = 6 \pi^2 {\bar\lambda} \left(\frac{\mpl}{M}\right)^4
\,,
\ee
where we have defined
\be
H_0^2 \equiv \frac{V_0}{3 M_{pl}^2} = \frac{1}{3} \frac{M^4}{\mpl^2}
\,.
\ee

For the above compactification data, $N = 60$, giving a 
BM bound (\ref{bmbound}) of $\Delta\vp_{max} = .26$.  
In our examples we take the initial value of the inflaton to 
be well within this limit,  $\vp_i = .15$ and near the inflection point.

\begin{figure}[htbp!]
    \centering
        \includegraphics[width=0.45\textwidth]{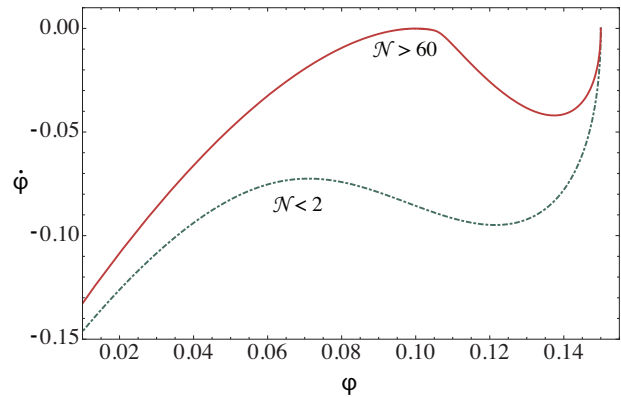}
    \caption{Phase portrait showing the inflaton trajectory in the inflection point potential for (a) minimally coupled inflaton (solid red) and
    conformally coupled inflaton (dashed gray). }
    \label{etap.eps}
\end{figure}

Using the above data it is possible to build a slow-roll inflationary 
model with $\mathcal{N}\geq 60$ e--foldings near the inflection 
point of (\ref{inpot}) 
if the inflaton is minimally coupled ($\xi =0$).
We provide a corresponding numerical solution in 
Fig.~\ref{etap.eps}. The solution is an
example of a successful slow-roll inflationary solution 
discussed in \cite{Baumann:2008kq}. 

The situation changes dramatically, however, if we include the effects of the
conformal ($\xi =1/6$) coupling term in (\ref{actst}). 
Before discussing the altered solution, let us take a moment 
to ascertain some of the general consequences
of including the  $\xi$-term.  First, the mass term introduces 
an effective Planck Mass:
\be
M_{\rm eff}^2  \equiv  \mpl^2 - \xi \vp^2
\,.
\ee
For a sensical gravitational theory we demand this remain positive, 
$M_{\rm eff}^2  >0$ \cite{Starob}. For the case $\xi>0$, this 
introduces an effective UV cutoff for
the theory, at the critical value
\be\label{cbound}
\frac{\vp_c}{\mpl} = \frac{1}{\sqrt{\xi}}
\,.
\ee
Thus, the introduction of the $\xi$-term can destroy 
large field models of inflation in general.
This cutoff exists independently from the field range bound (\ref{bmbound}). In fact, the bound 
(\ref{cbound}) is more general than the BM bound which holds 
only in the specific instance of the warped throat brane inflation
scenario. In a typical setting, $N> 4 \xi$, so that the BM bound, 
if applicable, is the more restrictive of the two. Finally, we mention
this cutoff will have strong implications for models of inflation
where $\vp$ is growing, and hence, approaching the critical
value. Such is the case in infra-red (IR) models of brane inflation~\cite{Chen:2005ad}.

Secondly, we can think of the conformal coupling term
as generating an effective potential (see (\ref{nreomp})):
\be\label{veff}
V'_{\rm eff}(\vp) = \vp (2 H^2 + \dot H) + V' 
\,,
\ee
effectively  steepening the potential, decreasing the possibility for 
slow-roll inflation. This effect has noteworthy implications
for the existence of eternal inflation \cite{Vilenkin:1983xq} in the brane anti-brane model
of~\cite{KKLMMT}, or other models with very flat potentials. 
Regions of the
Universe where eternal inflation can occur have an indefinitely large and growing volume, 
making these regions statistically favored~\cite{Linde:1993xx}. 
For inflating solutions with very flat potentials,  $\dot \vp \simeq \dot H \simeq V' \simeq 0$, (\ref{nreomp})
implies the conformal term generates a new downward force $\ddot \vp \simeq  - 2 \, H^2 \vp$
typically decreasing the number of favorable regions where eternal inflation can occur.

In the numerical analysis plotted in Fig.~\ref{etap.eps} 
we see that including the conformally coupled scalar allows for 
only $\mathcal{N} \simeq 2$ e--foldings of inflation (although there 
may be an interesting possibility of rapid-roll inflation at 
sufficiently low inflationary scales, see \cite{Kofman:2007tr}).

\begin{figure}[htbp!]
\centering
\includegraphics[width=0.45\textwidth]{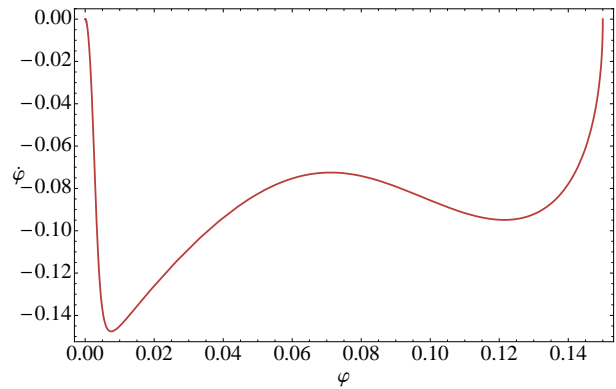}
\caption{Phase portrait for inflaton with DBI kinetic term and $\xi =1/6$.  
$\mathcal{N} \geq 60$ e--foldings are achieved.}
\label{dbi.eps}
\end{figure}

Using the same compactification data and initial conditions as above, 
we now appropriately take into account the non-standard DBI form of 
the kinetic term for the inflaton field. Thus, the entire action for 
the system is (\ref{actst}) added to
\be\label{dbigrav}
\frac{\mathcal{L}_{DBI}}{ \sqrt{-g} } = 
- \frac{ 1}{f(\vp)} \left[ \sqrt{ 1 + f(\vp) \, g^{\m\n}\, 
\partial_{\m}\vp \, \partial_{\n} \vp} -1 \right]
\,,
\ee
and potential (\ref{inpot}).
\begin{widetext}
Variation with respect to the metric $g^{\mu\nu}$ leads to:
\be
(M_{pl}^2 - \xi \vp^2) G_{\m\n}  =
 \frac{\d_\m \vp \, \d_\n \vp -  g_{\m\n} \,\left(\d^\s\vp \,\d_\s \vp + f^{-1}(\vp)\right)}{\sqrt{1 + f (\vp)(\d \vp)^2}}
- \xi \Big(\nabla_\m \nabla_\n (\vp^2) - \nabla_\s  \nabla^\s (\vp^2) g_{\m\n}\Big) + \left(f^{-1}(\vp) - V(\vp) \right) g_{\m\n} 
\,.
\ee
The equation of motion for the field $\vp$ is: 
\be
\nabla_\m ( \gamma \, \d^\m \vp ) + f^{-2} f' (\gamma^{-1} - 1) \\ \nonumber 
 -\frac{1}{2} f^{-1} f' \, \gamma\, (\d \vp)^2 -  V' -  \xi R \vp =0
\,,
\ee
\end{widetext}
%\bea
%\nabla_\m ( \gamma \, \d^\m \vp ) &+& f^{-2} f' (\gamma^{-1} - 1) \\ \nonumber 
% &-& \frac{1}{2} f^{-1} f' \, \gamma\, (\d \vp)^2 -  V' -  \xi R \vp =0
%\,,
%\eea
where
\be\label{gamma}
\gamma \equiv \frac{1}{\sqrt{1 + f(\vp) (\d \vp)^2}}
\,,
\ee
and prime denotes differentiation with respect to $\vp$.
The resulting Friedmann equation and equation of motion for $\vp$ are:
%\bea\label{eomn}
%3 \left(M_{pl}^2 - \frac{\vp^2}{6} \right) H^2 
%=  f^{-1} \left(\frac{1}{\sqrt{1 - f \dot \vp^2}} -1\right) \nonumber \\
%+  H  \vp \dot \vp  + V(\vp)
%\,,\\
%\frac{\ddot \vp}{(1 - f \dot \vp^2)^{3/2}} 
%+ \frac{3 H \dot \vp}{\sqrt{1 - f \dot \vp^2}} + V'  \nonumber \\
%+ \frac{f'}{f^2} \left( \frac{3f \dot \vp^2 -2}{2 (1 - f \dot \vp^2)^{3/2}} 
%+ 1 \right)+  \vp (2 H^2 + \dot H) =0
%\,.
%\eea
\bea\label{eomn}
3 \left(M_{pl}^2 - \frac{\vp^2}{6} \right) H^2 
&=&  f^{-1} \left(\gamma -1\right)
+  H  \vp \dot \vp  + V(\vp)
\,,\\
\gamma^3 {\ddot \vp} 
+ 3 H \gamma {\dot \vp} &+& V'  
+  \vp (2 H^2 + \dot H) 
\nonumber \\
&+& \frac{f'}{2f^2} \left( (3f {\dot \vp}^2 -2) \gamma^3 
+ 2 \right)  =0 \,,\;\;\;\;\;\;
\eea
respectively, where $\gamma = (1-f{\dot\vp}^2)^{-1/2}$.
Solving the equations we find that taking into account the 
nonstandard nature of the kinetic term restores a successful 
inflationary model leading to more than 60 e--foldings of 
relativistic DBI inflation, circumventing the eta problem (Fig.~\ref{dbi.eps}).
While this
result is not entirely unexpected it has not yet explicitly been shown
in the literature in the context of a conformally coupled inflaton.
\section{Observational Predictions}
As we have already mentioned, it is not our goal to fit the above 
model to the latest observational data, but rather to present
a solution that avoids the eta problem and successfully produces 
more than 60 e--foldings of inflation while simultaneously satisfying
all known compactification bounds. However, despite our current
modest goals, it is interesting to understand in generality the possible effects of
gravitational couplings in theories with nonstandard kinetic terms  on cosmological observables
and we initiate this study here.

In this Section we show that the presence of a non-minmally coupled 
inflaton field in a DBI action changes the overall behavior of the sound
speed in the model. In principle, this can lead to significant observational effects,
for example, in the form of non-gaussianities in the CMB, allowing for possible observational 
distinction between pure DBI and DBI with non-minimally coupled inflaton\footnote{CMB constraints on kinetically
modified inflation models were considered in, e.g.,~\cite{Lorenz:2008je}.}.

To calculate the observable quantities from the action (\ref{actst}) coupled to
(\ref{dbigrav}) we first perform the weyl rescaling:
\be\label{ctrans}
\tilde g_{\m\n} = \Omega^2(\vp) g_{\m\n}
\,,
\ee
where $\Omega^2$ is the scaling factor given in terms of the inflaton as
\be
\Omega^2 = 1 - \xi \frac{\vp^2}{\mpl^2}
\,.
\ee

Under the transformation the action (\ref{dbigrav}) becomes
\be\label{actp}
S = \int d^{4} \! x \,  \sqrt{- \tilde g} \, \left(\frac{ \mpl^2}{2} \tilde R + P(X,\vp) \right)
\,,
\ee
where 
$X \equiv - \frac{1}{2} \tilde g^{\m\n} \d_\m \vp \d_\n \vp$. The functional form for $P$ is given by:
\bea\label{xip}
\left(1-\xi \frac{\vp^2}{\mpl^2} \right)^2 \, P(\vp,X) = 6 \xi^2 X \frac{\vp^2}{\mpl^2} && \\ \nonumber
- f^{-1}(\vp) \sqrt{1 - 2Xf \left(1-\xi \frac{\vp^2}{\mpl^2} \right)} &+& f^{-1}(\vp) - V(\vp)
\,.
\eea

We denote this frame as the \it P-frame\rm. Note that this \it is not \rm the traditional Einstein frame,
since in the low energy limit the action for $\vp$ does not reduce to that of a cannonically normalized scalar field as
it does in the special case where $P$ is that of ordinary DBI inflation. We now calculate the physical 
observables directly in the $P$ frame \cite{Makino:1991sg}.
The theory of perturbations for actions of the form (\ref{actp}) was carried out in \cite{Garriga:1999vw}.
We will follow the study of primordial scalar non-gaussiantities in \cite{Chen:2006nt}.

The inflaton energy is given by:
\be
E = 2 X P_{,X} - P
\,,
\ee
and quantities of measurable interest are conveniently written in terms of the generalized slow-roll parameters:
\bea
\epsilon &=& - \frac{\dot{H}}{H^2} = \frac{X P_{,X}}{\mpl^2 H^2} \,,\\ \nonumber
\eta &=& \frac{\dot \epsilon}{\epsilon H} \,, \\ \nonumber
s &=& \frac{\dot c_s}{c_s H}
\,,
\eea
where $c_s$ is the sound speed:
  \be
  c_s^2 =  \frac{dP}{dE} = \frac{P_{,X}}{P_{,X} + 2 X P_{,XX}}
  \,,
\ee
which is most conveniently calculated in terms of the rescaled warp factor:
\be
\tilde f(\vp) \equiv f(\vp) \left( 1 - \xi \frac{\vp^2}{\mpl^2} \right)
\,.
\ee
The result for our $P(X,\, \vp)$ (\ref{xip}) is:
\be\label{cseq}
c_s^2 = \frac{(1 - \tilde f \dot \vp^2) ( 1 - \xi  \frac{\vp^2}{\mpl^2}  
+ 6 \xi^2 \sqrt{1 -  \tilde f  \dot \vp^2} 
\frac{\vp^2}{\mpl^2})}{  1 - \xi  \frac{\vp^2}{\mpl^2}  
+ 6 \xi^2 (1 -  \tilde f  \dot \vp^2)^{3/2}  \frac{\vp^2}{\mpl^2}}
\,.
\ee

In the limit of non-minimal coupling, $\xi =0$, (\ref{cseq}) reduces 
to the standard result for DBI inflation: 
\be\label{dbics}
c_{s,\, DBI}^2 = 1 - f \dot\vp^2
\,.
\ee
However, for the case under consideration, the results are slightly modified 
by the scalar-gravitational coupling term, allowing, at least in principle, for observationally distinct signatures.
The observable quantities of interest include, the primordial power spectrum:
\be
P^\z_k = \frac{1}{8 \pi^2 \mpl^2} \frac{H^2}{c_s \epsilon}
\,,
\ee
the tensor perturbation spectrum:
\be
P^h_k = \frac{2}{3 \pi^2} \frac{E}{ \mpl^4}
\,,
\ee
and their respective spectral indices $n_s$ and $n_T$.
%\be
%n_s -1 = -2 \epsilon - \eta - s \,, \qquad n_T = - 2 \epsilon
%\,.
%\ee
%The tensor to scalar ratio is given by:
%\be
%r = -8 c_s n_T
%\,.
%\ee
In this paper, we choose to focus on observational signatures in the form of deviations from 
gaussianity of the CMB. Non-gaussianities are sensitive to the three point correlation function
of the Fourier transform of the gauge invariant curvature perturbation $\z$ \cite{Mukhanov:1990me}. A useful way to quantify
the level of non-gaussianity in a given model is in terms of the scalar quantity, $f_{NL}$, given by
\be
\z = \z_L - \frac{3}{5} f_{NL} \z_L^2
\,,
\ee
where $\z_L$ is the linear gaussian part of the perturbation\footnote{Here we use the sign conventions for
$f_{NL}$ of \cite{Chen:2006nt}.}. Single field, slow-roll models
generically predict undetectable levels of non-gaussianity \cite{Maldacena:2002vr}. 
This precise form for $f_{NL}$ is valid for the \it local \rm form of non-gaussianity.  For the
$P$-frame action (\ref{actp}):
\be\label{fnl}
f_{NL} = \frac{5}{81} \left(1 - \frac{1}{c_s^2}  + 2 \Lambda \right) - \frac{35}{108} \left( 1- \frac{1}{c_s^2} \right)
\,,
\ee
where 
\be
\Lambda \equiv \frac{X^2 P_{,XX} + \frac{2}{3} X^3 P_{,XXX}}{X P_{,X} + 2 X^2  P_{,XX} }
\,.
\ee
For the case of pure DBI, with sound speed (\ref{dbics}), the first term in (\ref{fnl}) vanishes \cite{AST,nongauss}.
However, in the case of DBI plus non-minimally coupled scalar this is no longer the case 
(the second term is also altered with respect to the pure DBI form via the sound speed modification).
The overall change in the sound speed due to the presence of the non-minimal coupling term
opens up the possibility that non-gaussianity and other observables listed above may be used to probe gravitational
couplings of the form (\ref{conft}) (or other such terms). While we have focused on the DBI case above,
we expect similar behavior in any model that contributes measurable non-gaussianities such 
as the K-inflation models of~\cite{Garriga:1999vw}. 
\section{Conclusions}
In this paper we have explicitly demonstrated the ability of the DBI mechanism to 
circumvent the eta problem in a class of string inflation models. We have worked within
the context of a non-minimally coupled inflaton field, however, our results
apply to more general situations of pure Hubble mass corrections to the inflaton potential.
Sources of Hubble mass corrections, both in the context of effective field theory and
in the context of string theory were discussed in Section \ref{sec:pmt}.

DBI inflation models typically face a particular challenge 
due to backreaction effects and we have nothing new to add 
to this discussion here~\cite{McASilv,spin,dbibr}. 
Indeed, the sound speed in the above solution grows 
very small rapidly and produces large non-gaussianity in excess 
of current observational bounds. Nevertheless, our goal was to 
obtain an inflationary solution
that makes progress towards realizing inflation even with Hubble mass inflaton.
In the present context it is possible to construct tuned inflaton trajectories that are initially slow-roll and 
fall into the DBI regime only towards the end of inflation. Because of this, 
$f_{NL}$ can become large after the CMB scales set in. This leaves the possibility 
of a model gaining the majority of e--foldings from standard slow-roll and the rest from DBI 
with detectable, yet observationally acceptable, levels of non-gaussianity. We have not determined wether such solutions
are compatible with the stringy compactification constraints of~\cite{BM}.

Finally, we have pointed out that, because adding a non-minimally coupled scalar to the DBI or K-inflation
actions changes the formula for the speed of sound (and observable quantities), gravitational couplings can contribute
potentially observable signatures in the CMB. We leave a more detailed analysis of
this interesting possibility for future work \cite{Easson:2009wc}.
%\vskip 3mm 
%\pagebreak
%\noindent
\acknowledgments
%\begin{acknowledgments}
%{\it Acknowledgements.}\
It is a pleasure to thank Simeon Hellerman, Takeshi Kobayashi, Lev Kofman, Andrei Linde,
Juan Maldacena, Shinji Mukhoyama, Hirosi Ooguri, Brian Powell, Simon Ross, Henry Tye, Taizan Watari and Ivonne Zavala 
for helpful conversations. This work is supported in part by STFC.
DE is also supported by the World Premier International Research 
Center Initiative (WPI Initiative), MEXT, Japan and by a Grant-in-Aid for Scientific Research 
(21740167) from the Japan Society for Promotion of Science (JSPS) and by funds from the Arizona State University Foundation.
%\end{acknowledgments}

%\begingroup\raggedright
%\endgroup

\end{document}